\documentstyle[prb,twocolumn,aps]{revtex}

\begin{document}
\baselineskip=12pt

\draft
\flushbottom
\twocolumn[\hsize\textwidth\columnwidth\hsize
\csname@twocolumnfalse\endcsname
\title{Hall effect in the perovskite manganites}
\author{Pinaki Majumdar, Steven H. Simon and
Anirvan M. Sengupta}
\address{ Bell Laboratories, Lucent Technologies,
600 Mountain Avenue, Murray Hill, NJ 07974}

\date{\today}

\maketitle
\tightenlines
\widetext
\advance\leftskip by 57pt
\advance\rightskip by 57pt

We compute the zero temperature phase diagram and the Hall
response of the doped perovskite manganites within the model 
of double exchange and Jahn-Teller coupling, employing the 
$d\rightarrow \infty$ approximation proposed by Millis 
{\it et al.}. We find that in this scenario a ``hole - like'' 
$R_H$ for the hole doped manganites, as observed at low 
temperature by Matl {\it et al.} \cite{ong}, can be obtained 
only when lattice distortions persist in the {\it metallic state} 
as $T \rightarrow 0$. The calculated temperature dependence of
$R_H$ seems to be consistent with the measured ``normal'' 
Hall coefficient in thin films.

\

\


]

\narrowtext
\tightenlines


\section{ Introduction }

With the current interest in ``colossal magnetoresistance'' 
the ferromagnetic manganese oxides have been intensely studied
over the last few years \cite{cmrrev}. These compounds, of
the form 
 $La_{1-x}(Sr,Ca)_xMnO_3$, are antiferromagnetic insulators in the
undoped phase $(x=0)$, and have a ferromagnetic ground state
beyond a critical doping. The ferromagnetism is conventionally
understood in terms of the double exchange model 
\cite{doublex,furukawa}.
 The 
transition  from the ferromagnetic to the paramagnetic phase is 
accompanied by a simultaneous metal-insulator `transition'. The 
transport is typically activated in the paramagnetic phase.

The large increase in resistivity near  $T_c$ and the rather low
transition temperature prompted the suggestion \cite{millis1}
that, in addition to the double exchange mechanism,
carrier localisation due to 
electron-phonon coupling effects could be relevant 
in these materials.  Detailed calculations \cite{millis2,lanl} 
incorporating both double exchange and the   
Jahn-Teller coupling have been successful in
reproducing most of the qualitative
features of the data. So, although  there are possibly
complications due to  electron
correlation effects as well as disorder due to cation
substitution, the basic interplay between charge, spin and
lattice degrees of freedom in the manganites 
seems to be adequately
described by the double exchange/Jahn-Teller scenario.

Experimentally there is overwhelming evidence 
\cite{latticeff} for
dynamical  lattice distortions
in  these compounds, including  isotope effects on $T_c$ and
direct measurement of the Debye-Waller factor. 
 There is also a systematic and large change in lattice
constant with temperature.
 The activated transport in the
paramagnetic phase has been interpreted in detail in terms
of small polaron hopping. While it is obvious and
acknowledged that the electron phonon coupling is central to
understanding the ``high temperature'', $T \gtrsim T_c$, physics,
its effect on the low temperature metallic phase  is much less
studied. This spin polarised phase is believed to admit
an almost ``band theoretic'' description, with 
electron-magnon interactions
describing the transport \cite{jaimecmat}.
The measurement of the 
Hall effect \cite{ong} in  
 $La_{0.67}Ca_{0.33}MnO_3$ at low temperature, however, suggests
that the electron-phonon coupling may be crucial in
understanding this phase too, as we discuss next.

\section{ Hall measurements on the
manganites.}

The Hall response of the manganites has signatures of various
interaction effects. There is strong field dependence to the
Hall resistivity, as is usual in ferromagnets, and the data
can be parametrised as $\rho_{xy}= R_H(T) B + R_S(T) M$,
where $R_H$ and $R_S$ are respectively the ``normal'' and
``anomalous'' Hall coefficients, $B$ the induction and $M$
the spontaneous or induced magnetisation. 
Measurements indicate that  $R_H > 0$, while the anomalous
 term $R_S < 0$
(in the Ca doped system \cite{ong})
and $\vert R_S \vert \gg R_H$ for $T \sim T_c$ \cite{jaime}. 
In fact, at all but the lowest temperatures, $\rho_{xy}$
is dominated by the anomalous term. The sign and temperature
dependence of $R_S$ is intriguing and it has been suggested
that its origin may lie in a combination of 
spin-orbit coupling effects 
and strong electronic correlations \cite{ybkim}. Here we 
concentrate on the normal
Hall coefficient which, although much smaller than $R_S$,
also has several mysterious features, not necessarily
linked to the ferromagnetism; $(a).$ $R_H$ is ``hole-like'' 
although a simple electron count with degenerate Mn  orbitals
would predict an ``electron-like'' Hall coefficient.
 $(b).$ The carrier
density, inferred from $R_H^{-1}$, at $x=0.33$ seems to be
five times the nominal doping.
So, either the Drude form   $R_H^{-1} \sim x$, naively 
expected for the 
doped insulator, is inapplicable,
or there is a serious discrepancy between the nominal and
actual carrier density. Therefore, beyond the 
anomalies attributable to
spin orbit coupling effects, the doping and
temperature dependence of the normal Hall coefficient itself  in
the manganites is not understood.
We have calculated $R_H$ within the model of 
double-exchange and  Jahn-Teller
coupling  introduced by Millis {\it et al.} as a function of
electron-phonon coupling, $g$, electron density, $n$, and
temperature, $T$. 

\section{ Model}

We use the standard Hamiltonan,
$$
{\hat H}= 
{\hat H_{el}} + {\hat H_{d-ex}} + {\hat H_{ph}} +{\hat H_{el-ph}}
$$
The electronic band is specified by 
${\hat H_{el}} = \sum_{ij}
t^{\alpha \beta}_{ij}c^{\dagger}_{i\alpha \sigma}  
c_{j \beta \sigma} $ - $\sum_i \mu n_i$, where $\alpha$ and 
$ \beta$ 
refer to the two $e_g$
orbitals.
The double-exchange term
${\hat H_{d-ex}} = 
-J_H\sum_i {\vec S}_c^i . {\vec \sigma}_i$ couples the conduction
electrons 
to the core `spin' $3/2$, which we
treat  classically,
assuming  $J_HS_c/t \rightarrow \infty$.
The classical phonons are described by a
harmonic distortion term ${\hat H_{ph}}=
\sum_i kr_i^2/2$, and the Jahn-Teller coupling
${\hat H_{el-ph}}= g\sum_i {\vec r_i} .
c^{\dagger}_{i \alpha\sigma}{\vec \tau_{
\alpha \beta}}c_{i\beta \sigma}$
where ${\vec \tau}$ is ${\hat x} \tau_x + {\hat y} \tau_y$,
 and $\tau$'s are Pauli matrices.
This model, even with spins and phonons 
treated classically, can be solved exactly
only in the $d\rightarrow \infty$ limit. We follow
the prescription of 
Millis {\it et al.}
\cite{millisprb}  in handling the problem
and the reader should consult their papers for the 
 detailed 
formulation within the $d \rightarrow \infty$ framework.
 We skip the
technical details here, referring the interested reader 
to our appendix for a discussion of the
$T=0$ limit of the equations.

For following our results  it is sufficient to know
that this scheme provides us with the 
distribution of lattice distortions and
local spin orientation, $P(r, \theta)$, 
and the electronic Greens functions
$G_{\sigma \sigma} ({\vec k}, \omega)$,
for electrons with spin parallel/anti parallel
to the magnetisation.
The transport coefficients are computed from the spectral
functions, $A_{\sigma \sigma}
({\vec k}, \omega)= -(1/\pi) Im G_{\sigma \sigma}
({\vec k},\omega)$ as
\cite{dinftransp},
$
\sigma_{xx} = 
\sigma_{xx}^{\uparrow \uparrow}
+ \sigma_{xx}^{\downarrow \downarrow} 
$ and $\sigma_{xy}=  
\sigma_{xy}^{\uparrow \uparrow}
+ \sigma_{xy}^{\downarrow \downarrow}
$
where, for example, 
$\sigma_{xx}^{\uparrow \uparrow} = \int d\epsilon
 \rho_0(\epsilon) 
 \int d\omega
(-\partial f/{\partial \omega})
A_{\uparrow \uparrow}^2(\epsilon_{\vec k}, \omega)$,
and
$\sigma_{xy}^{\uparrow \uparrow} = \int d\epsilon
 \rho_0(\epsilon) \epsilon
 \int d\omega
(-\partial f/{\partial \omega})
A_{\uparrow \uparrow}^3(\epsilon_{\vec k}, \omega)$, 
$\rho_0(\epsilon)$ is the free band density of states,
and $f$ is the Fermi function.
The low field Hall coefficient, as usual, is $R_H = \sigma_{xy}/
\sigma_{xx}^2$. We work in units where the half-bandwidth,
$D=2$. 

Unlike in [12]
where most of the calculations were done at particle-hole symmetry,
we have to  work away from $\langle n_i \rangle =1$ to obtain
a non trivial Hall coefficient. We have solved the
self-consistent equations numerically  for various $g, n $
and $T$ and
discuss the results next.

\section{Results}

\subsection{ $T =0$, phase diagram, and doping  dependence
of $R_H$}

At low temperature,  $T \ll T_c$, 
the spin background is  almost completely polarised.
In the absence of any orbital ordering, the phonon 
distribution depends only on the scalar distortion,
$r= \vert {\vec r} \vert$,
and, for $T \rightarrow 0$,
the  probability distribution, $P(r)$, for
the (classical) phonons
is sharply peaked, since the variance $\langle  {\Delta r}^2 
\rangle \sim T$.
Therefore, in this limit, 
the electronic spectrum  is completely determined
by the average onsite distortion 
$\langle  r \rangle =   r_0$.
This permits a simple classification of
the various regimes as $T \rightarrow 0$. Following  
[12]  these are
$(a)$. Fermi liquid;
$\langle r \rangle \rightarrow 0$
as $T \rightarrow 0$, 
electron-phonon interaction effects vanish 
and there are no ``quenched'' distortions.  
$(b)$. ``Rigid band'' phase, $0 <  {\langle r \rangle}
<   {r_c(n)}$, where 
$ {r_c(n)}$ denotes the critical distortion beyond which
the system, with electron density $n$, becomes a
polaronic insulator, and $(c)$.
 $\langle r \rangle \ge  {r_c(n)}$
corresponding to the insulating state.
Obviously, since phononic effects completely disappear in the
first case, the Hall effect would simply correspond to that
of a spin polarised band, {\it i.e} the band theoretic result.
In the third case, 
since the conductivity vanishes as $T\rightarrow 0$,
the Hall coefficient is ill-defined. The most interesting regime 
corresponds to $(b)$, where lattice distortions persist in the
$T\rightarrow 0$ metallic state and $R_H$ can significantly differ,
in sign and magnitude, from the band-theoretic value.
The metallic or insulating character of the ground state is
determined from the density of states, $N(\omega )$, at
$\omega =0$. For the Fermi liquid this corresponds to
the band DOS, for the `rigid band' phase it is
finite but suppressed from the band value, and
for the polaronic insulator $N(0)=0.$

Our result for 
the ``phase boundaries'' separating the Fermi liquid,
the distorted metallic phase and the polaronic insulator at
$T=0$
are shown in Fig.1. 
Before discussing the behaviour of  
$R_H$ in the various phases let us establish the band
theoretic result. Consider half-filling first, ignoring the
antiferromagnetism for the moment. If the two $e_g$ orbitals
were equally occupied, and the spin states were degenerate,
then the ``half-filled'' state would
correspond to quarter filling of spin degenerate
bands, and the
Hall effect would be ``electron-like''. However, if 
the $T\rightarrow 0$
state were spin polarised the majority carriers would 
inhabit  two
orbitally degenerate {\it half-filled} bands. This state, for
a particle-hole symmetric bandstructure would correspond to
$R_H =0$. 
Hole doping on this state, {\it i.e} a reduction
of the electron count, {\it would lead to an  electron-like
Hall effect, not hole-like as has been measured}. 

The numerically obtained $T\rightarrow 0$ result for $R_H$ confirms
that for  $g < g_c(n=1) \sim 1.14$, where the half-filled state becomes
insulating, the Hall coefficient, at all doping, correspond to
the band prediction. These states all have vanishing distortion
as $T\rightarrow 0$. The results on $R_H$, for several
  interaction
strengths, $g=0.9,1.0, 1.3$, 
 are superposed in Fig.2.  Between  $g_c(1)$ and $ g'_c(1)$
the half-filled state remains  insulating 
while  states away from
half-filling are metallic, but with finite $\langle r \rangle$.
However, for a given $g$, with $g_c(1) < g < g'_c(1)$,
beyond a critical deviation away from half-filling the
distortion seems to disappear. In this region, bounded
by $g_c(n)$ separating the Fermi liquid and distorted phase,
$R_H$ exhibits interesting deviation  from the 
band value. Notice that while $R_H=0$ would occur only at the
particle-hole symmetric point, $n=1$, in the Fermi liquid
phase, it occurs {\it away from half-filling} in the 
distorted phase. This is the key to the anomalous behaviour in 
the doped insulator.
 
 We have shown a typical plot, for $g=1.3$,  in 
Fig.2,  to
illustrate this behaviour.
Notice that with approach to half-filling, where the interaction
effects are strongest due to the lowest kinetic energy, $R_H$
moves off the band result, {\it changes sign somewhere}, and
tends to diverge as $n \rightarrow 1$.
It is numerically difficult to push the imaginary 
frequency calculation below $T \sim 0.01$, but an
asymptotic analytic  solution (see Appendix) at $T=0$
indicates that for $x\rightarrow 0$, $R_H \sim x^{-1/3}$.
Increasing $g$ beyond $g'_c(1)$ states away from half-filling
rapidly turn insulating. The boundary $g'_c(n)$ separating
the polaronic insulator from the Fermi liquid and
distorted metallic phases is shown in Fig.1. 

\subsection{Temperature dependence of $R_H$}

Electron-phonon coupling leads to a suppression of
electron kinetic energy, as does magnetic disorder
within the double exchange model. So, qualitatively,
the effect of increasing temperature, reducing
magnetisation, and suppressing
electron motion, would look like increasing $g$ at
$T =0$.  The finite $T$ results bear this
out, as shown in 
Fig. 3. For the chosen coupling constant, $g =1.5$, 
for $x=0.2$ the hole-like $R_H$ at $T=0$ becomes
larger as $T \rightarrow T_c$,  for $x=0.3$ the
low temperature {\it negative} Hall coefficient
changes sign to hole-like at some intermediate
temperature, while for $x=0.4$ it becomes less
negative with increasing $T$ but never changes sign. 
A complementary situation would hold for `electron-doped'
systems, if they could be experimentally realised.
Overall, $R_H$ deviates even more from its band
theoretic value with increasing temperature $(T \le T_c)$.

Within our approximation the magnetic disorder is 
maximal at $T=T_c$ (the spin distribution 
becomes isotropic)
and there is no further suppression of kinetic energy for
$T > T_c$. In this regime the behaviour of $R_H$ is
probably determined by the activation effects in $\sigma_{xx}$.

\section{Discussion} 
A `hole-doped' insulator is intuitively 
expected to have a hole-like $R_H$, so  our approach 	
might seem to be an elaborate way of arriving at
an obvious conclusion. The state we describe is, however,
rather non trivial, and below we contrast it both 
to $(a)$ the
doping of a simple band insulator and $(b)$ the
mean field picture of the ``distorted''
metallic  phase (where every site has the same distortion
${\vec r}$). Next, $(c)$,  we contrast
  the properties of this mean-field state
to that with randomly oriented distortions and,
finally, $(d)$,  comment on quantum effects
which we have neglected in our calculation.

$(a).$ 
Usually ``electron'' or
 ``hole'' doping refers to the doped carriers
occupying a specific part of the Brillouin zone. For a 
band insulator, the doped holes, by definition, occupy
the hole like parts of the zone, and lead to a
positive $R_H$. When the insulating state arises out of
interaction effects, at an electron density where 
band theory would have predicted a metal, 
 it is not clear at all
that a reduction in electron density is equivalent to
``hole doping'' in the earlier sense. Also, if
$R_H$ as  $T \rightarrow 0$, 
 away from the insulating state, is
required to deviate from the simple band value, some signature
of interaction effects {\it must} persist in the metallic
ground state. We have stated this earlier in the context
of our results, here we would like to emphasize that,
quite independent of specific models, a change in the 
sign of the $R_H$ at $T=0$ requires ``quenched'' 
distortions of some sort.
Proximity to an 
insulator is necessary, but not sufficient, to
obtain the behaviour observed.

$(b).$
Since the  Jahn-Teller
coupling  can lead to a lattice distortion and splitting of
the electronic band, naive  band theory predictions based on
degenerate $e_g$ levels are incorrect, as we have
argued before. 
One could, however,
construct 
 a mean-field picture {\it incorporating
Jahn-Teller distortions}, which might well provide an
adequate description of the half-filled
 insulating state 
\cite{feiner}.
By continuity one might argue that a 
similar mean-field picture for  
Jahn-Teller distortions {\it away from
half-filling}, with a self-consistently computed ${\vec r}$
would provide a reasonable description of
the metallic phase. In that case one would indeed
dope into the `lower band', and the quasiparticles,
in this orbitally ordered phase, will show a
hole-like $R_H$.
 That situation, though 
conceptually appealing, 
does not seem to be experimentally
relevant. There is no signature for orbital long range order
in the metallic state of the manganites
and one is required to deal with spatially random
distortions, as we have done. 

$(c).$ The phase we consider  has some 
distinct, non mean-field like, properties. 
Within the scheme we have adopted for the transport calculation,
the single particle spectral function completely determines
the transport coefficients. Since the spectral function
$A({\vec k}, \omega) =-(1/\pi) Im 
\{ \omega + \mu -\epsilon_{\vec k} -
\Sigma(\omega) \}^{-1}$,
the low frequency behaviour 
$lim_{ \omega \rightarrow 0} 
\{ \mu - \Sigma(\omega)\} \sim
{\bar \mu} -i \pi  \Gamma$,
determines the low temperature transport.
Here
$\bar {\mu} =\mu -\Sigma_R(0)$ and
$\Gamma = -(1/\pi) \Sigma_I(0)$
These parameters, $\bar {\mu}$ and $\Gamma$ have
a physical interpretation. $\Gamma$ is the single
particle scattering rate, and determines 
 the resistivity
within this model, while 
$\epsilon_{\vec k} = {\bar \mu}$ determines  the
location of the  Fermi
surface and the Hall coefficient.
This distinction between the intuitive
mean-field picture and the orbitally disordered phase
shows up in the behaviour of these parameters, and 
have direct
physical consequences. The implications 
are $(i)$ When  the electrons propagate in a background
of random distortions, they will experience a large
scattering rate, in contrast to the orbitally ordered
phase where $\Gamma$ and 
$\rho(T) \rightarrow 0$ as $T \rightarrow 0$.
$(ii)$ The orbitally ordered metallic phase will show
a Fermi surface in photoemission, while the state
computed by us does not have any discontinuity in
its momentum distribution function since, near half-filling,
since one finds ${\bar \mu} > 2D$ the bandwidth
 (see Appendix).
It is somewhat difficult to cleanly verify these effects
in the electronic spectrum because there is already large
substitutional disorder, leading to a finite $\Gamma$ as
$T\rightarrow 0$, and a smearing of the Fermi surface.

$(d).$
We have  made a classical approximation
for the phonons and, even in the
absence of orbital ordering,
below some coherence scale the quantum
effects could become important and anneal out the
random lattice distortions. In this regard 
we can gain some 
insight from what is known about doped Mott
insulators.  In the Hubbard
model at large U, over an intermediate temperature
range, one can view the system as electrons interacting
with static 
random local moments. This doped Mott insulator has
a doping dependence of $R_H$ 
similar to what we have found here.
However, below $T \sim xD$, where $x$ is the doping,
a Fermi liquid ground state is recovered and $R_H$
becomes band like again. 
A similar situation could arise in this model
also, as we outline below.
Suppose  the relevant optical phonon
in our case has a frequency $\omega_0 = \sqrt {k/m}$. For 
$T \ll \hbar \omega_0/k_B$ we could integrate out the
phonons and generate a term  in the Hamiltonian of the form 
${ \lambda^2 \over { \hbar \omega_0} } \sum_i (c^{\dagger}_{i \alpha} 
{{ \vec \tau} \over 2}  c_{i\beta})^2.$ 
At $T =0$ the model reduces to a ``Hubbard'' model involving
two orbital species $c^{\dagger}_{i \uparrow \alpha}$ with
an effective onsite repulsion $U_{eff} = U_{coul} + 
{\lambda^2 \over { \hbar \omega_0}}$ (at finite $T$ we need to
consider the other spin species too since the ferromagnet is
not fully polarised). This suggests that there could be
a temperature scale below which
our results, involving static distortions,
 would not hold. 
 Apart from
the presence of substitutional disorder,
it is not clear why such a  quantum scale is 
not visible in the manganites.

We can now discuss how 
the measured $R_H$ fits within this
framework. Since experimentally the half-filled state is
insulating the relevant regime for us would be $g > g_c(1)$.
Let us choose $g=1.3$, as shown in Fig.2, for illustration.
There $R_H$ is indeed large and positive near
half-filling. {\it However}, the behaviour changes for
$x \sim 0.3$ where $R_H \sim 0$ and changes sign at larger
doping. This is effectively the interpolation between  
a
doped insulator (strong coupling, low kinetic energy)
and `band metal' (weak coupling, large kinetic 
energy). So the enhancement seen\cite{ong} in $R_H^{-1}$
at $x=0.33$ could be simply due to the proximity
of $R_H$ to a zero crossing. For our choice of $g$,
$x=0.33$ itself is electron like, but that depends
on the exact value of the coupling, which is anyway
unknown. The two predictions one can make within
this scheme are; (a). With increasing doping, within
a given chemical family, 
 $La_{1-x}Ca_xMnO_3 $ say, $R_H$ should change
sign from hole-like to electron like, unless
some other ordering phenomena intervenes
(b). With increasing {\it coupling},
 $Pb$ substitution on $Ca$, keeping
the doping level fixed, $R_H$ should deviate
further from its band-theoretic  value (should
become larger if it is already hole-like, etc).. 

On the $T$ dependence of $R_H$ itself, there does
not seem to be any published data. Unpublished
results from Bryan Lin {\it et al.} 
\cite{linhall} for
 $La_{0.67}Ca_{0.33}MnO_3 $  indicates that
the hole-like $R_H$ at $T=0$ increases with
$T$ and has a peak near $T_c$, beyond which
it seems to fall off. This seems to be
consistent with our calculated $T$ dependence
($x=0.2$, Fig.3). However, $\rho_H$ measurements
on the manganite are rather difficult, due to the
large magnetoresistance, and it would be useful
to have independent confirmation of this trend.

We would like to thank 
 Boris Shraiman, Y.B.Kim,
Art Ramirez and Harold Hwang for several
discussions.

\section{Appendix; The $T \rightarrow 0$ equations}

As we have seen,  much of the physics of $R_H(x,T)$ is 
controlled by the coupling constant and doping dependence {\it at
} $T=0$, so it may be helpful to consider the 
$T=0$ equations in some detail.
We essentially follow the derivation in [12].
Since the spins are completely polarised, it
is sufficient to consider only 
the Greens function for electron spin parallel 
to the magnetisation, $G_{\uparrow \uparrow}$.

Within the $d \rightarrow
\infty$ framework one solves a local
problem, of a site hybridising with a 
self-consistently computed bath, 
with the local Green's function $G^{loc}(\omega)$
being the local
projection, $G_{ii}(\omega)$,
 of the lattice Green's function. 
Broadly, for our electron-phonon problem,
there are three steps;
$(a).$ Compute the local Green's function, 
$G^{loc}_{\uparrow \uparrow}
( r,\omega)$ 
 for a 
 given value of local distortion, $r$.
This is elementary since the local fermion problem is
quadratic. 
$(b).$ Compute the 
 full local
Greens function by averaging over
coordinates, with the appropriate
distribution function.
$$
G^{loc}_{\uparrow \uparrow}(\omega)= \int d{\vec r}
P( r) 
G^{loc}_{\uparrow \uparrow}
(r,\omega)
$$
This is possible since the phonons are classical.
$(c).$ Calculate the distribution function, 
$ P(r)$, as the electronic free energy 
for  distortion $r$  as
$P \sim e^{- Tr ln G(r)}$.  This closes the
self consistent loop. 

More specifically, for the semicircular band DOS that
we have assumed, the following equations need to be 
solved for $T \rightarrow 0$. 
\begin{equation}
a(i\omega_n) = i\omega_n + \mu -
{1 \over {2}} \int d{\vec r}
P(r) \{ {1 \over { a - g r}}
+  {1 \over { a + g r}} \}
\end{equation}
which combines steps $(a)$ and $(b)$ above.
$a(\omega)$ is related to 
$G^{loc}_{\uparrow \uparrow}(\omega)$ by
$$
G^{loc}_{\uparrow \uparrow}(\omega)=
{1 \over {2}} \int d{\vec r}
P(r) \{ {1 \over { a - g r}}
+  {1 \over { a + g r}} \}
$$
The phonon distribution
\begin{equation}
P(r)= {1 \over {Z_{loc}} } exp[ -{r^2 \over {2T} }
+ \sum_n ln( { { a_n^2 -g^2r^2} \over {(i\omega_n)^2} } ) ]
\end{equation}
and $Z_{loc} = \int d {\vec r} P(r)$. 
A more complicated version of these equations hold at finite $T$,
where the distribution of local spin orientations also has to
be taken into account.

To understand the doping dependence of transport
coefficients we need to solve the equations above
in the case where
$P(r)$,
in the limit $T \rightarrow 0$, reduces to a 
delta function at one non zero value, {\it i.e},
\begin{equation}
a(i\omega_n) = i\omega_n + \mu -
{1 \over {2}} 
\{ {1 \over { a - g r}}
+  {1 \over { a + g r}} \}
\end{equation}
and $r$ is determined by maximising $P(r)$.
We can view this as a cubic equation for $a(\omega)$, 
for a specified $gr$.
It is possible to solve this equation near the
band edge, parametrised in terms of the deviation of the chemical 
potential $\mu$ from $\mu^*$ 
(where  one begins to  dope into the lower band), and the
DOS looks like,
$$
-(1/\pi) Im G^{loc}_{ret}(\omega) \sim \sqrt
{\omega - (\mu^* -\mu)} 
\theta
(\omega - (\mu^* -\mu))
$$
From this the number of doped holes can be estimated to
be $x = 1-n \sim (\mu^* - \mu)^{3/2}$.

To obtain $\sigma_{xx}$ etc, we need to use the
relation
$$
\mu - \Sigma(\omega) = G_{ret}(\omega) + G^{-1}_{ret}(\omega)
$$
which, for $Re G(0) \neq 1$, leads to $Im \Sigma(0)
\sim Im G(0) \sim (\mu^* - \mu)^{1/2}$. Combining
this with the result for $n$ before, we obtain
$\Gamma \sim x^{1/3}$. Also $\mu - Re \Sigma(0)
\sim Re G(0) + Re G^{-1}(0) \ge 2$ since $Im G(0) 
\rightarrow 0$. 

Since $\Gamma \rightarrow 0$ while $\mu - Re \Sigma(0) -
\epsilon_{\vec k} \neq 0$, the spectral functions, which
enter the transport calculation can be roughly
approximated as $A( {\vec k}, \omega) \sim \Gamma/
(\epsilon_{\vec k} - \bar \mu)^2$. From which,
employing the formulae cited previously in the text,
one obtains $\sigma_{xx} \rightarrow \Gamma^2$, {\it i.e}
$\rho \sim x^{-2/3}$, and $\sigma_{xy}
\sim \Gamma^3$. This leads to
$R_H \sim \sigma_{xy}/\sigma_{xx}^2 \sim x^{-1/3}$.

{\bf Figure Captions}

Fig.1:
Phase diagram at $T=0$ with varying electron density, $n$,
and Jahn-Teller coupling, $g$. The phase above the solid line,
$g'_c(n)$, is 
a polaronic insulator.
The shaded phase, bounded by $g_c(n)$ and 
$g'_c(n)$, has $\langle r \rangle \neq 0$
and is metallic {\it except at half-filling}
where it is insulating.
 The dashed line corresponds to $R_H =0$ which
occurs away from half-filling in the distorted metal.
In the region bounded by $R_H=0$ 
and $g'_c(n)$
$R_H$ has the `wrong' sign with respect to
band theory.

Fig.2:
Density dependence of $R_H$ (arb. units)
 at $T=0$ for
various J-T coupling. $g=0.9$ (circles),  $g=1.0$ (squares),
$g=1.3$ (triangles). The coincident $g=0.9$ and $1.0$ traces 
are the same as
the band theory result $(g=0)$.

Fig.3:
Temperature dependence of $R_H$ (arb. units) for
$g=1.5$ and different dopings, $x= 1-n$.
Results for $x=0.2$ (circles), $0.3$ (squares) and $0.4$
(triangles).

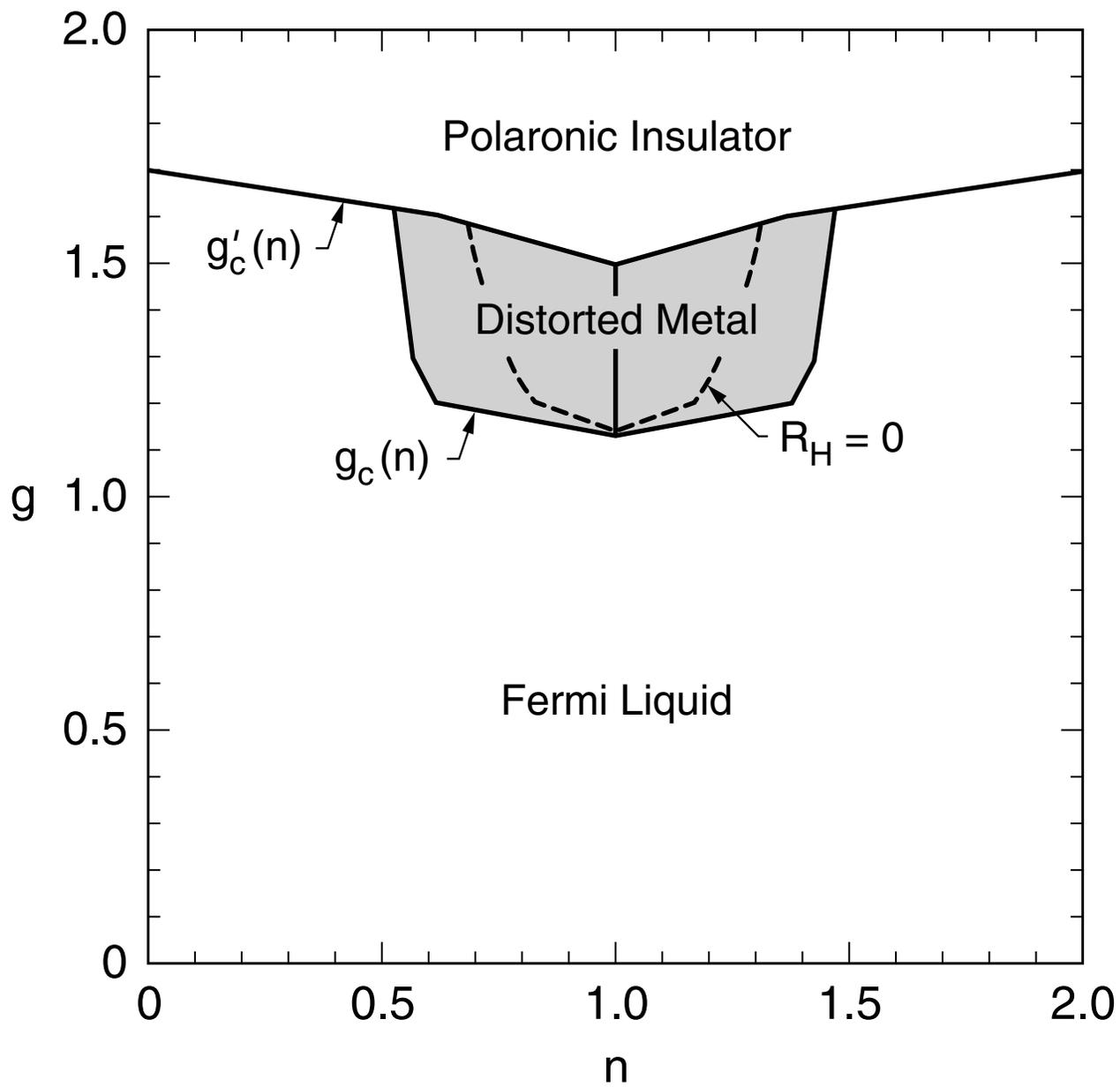
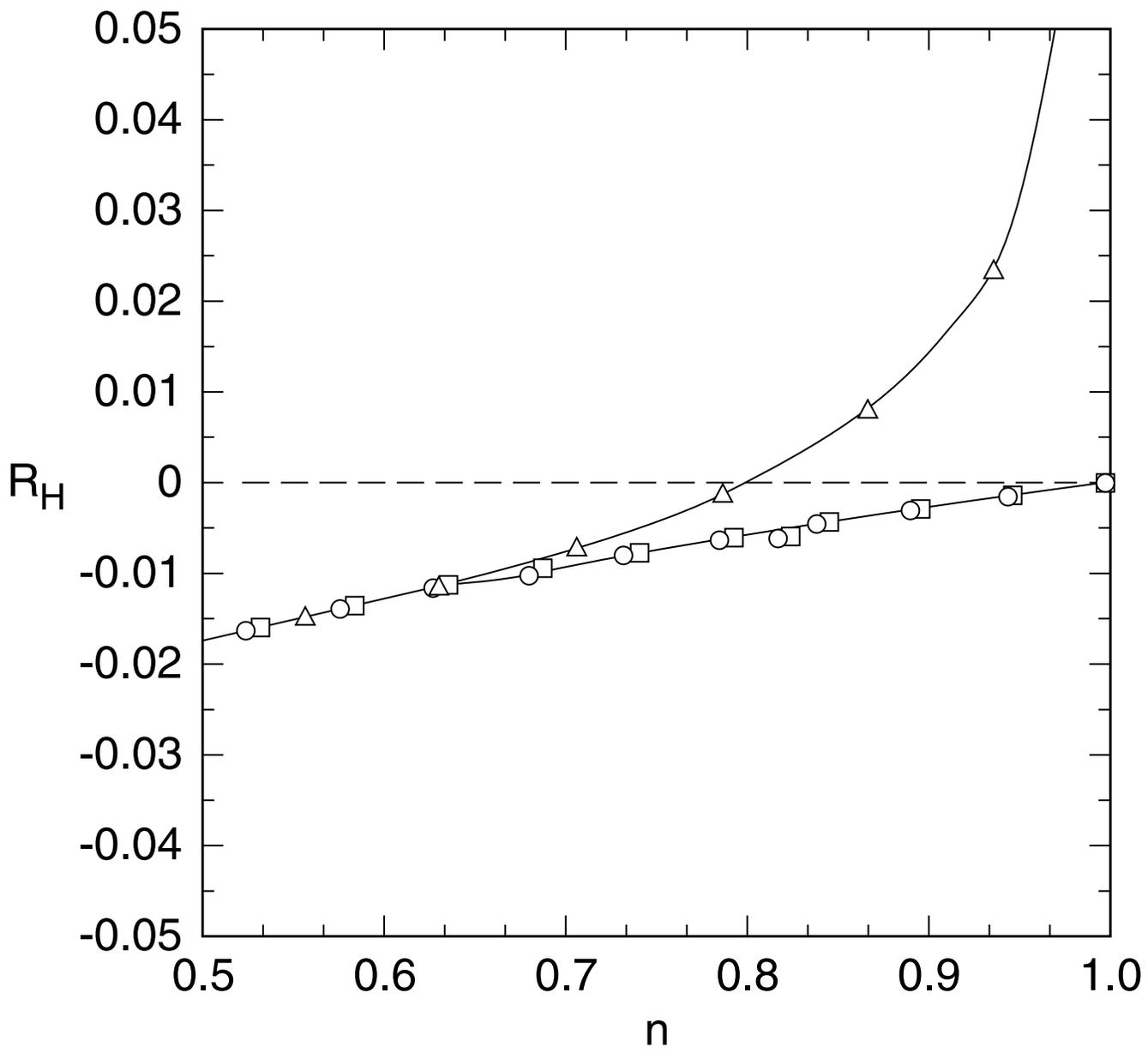
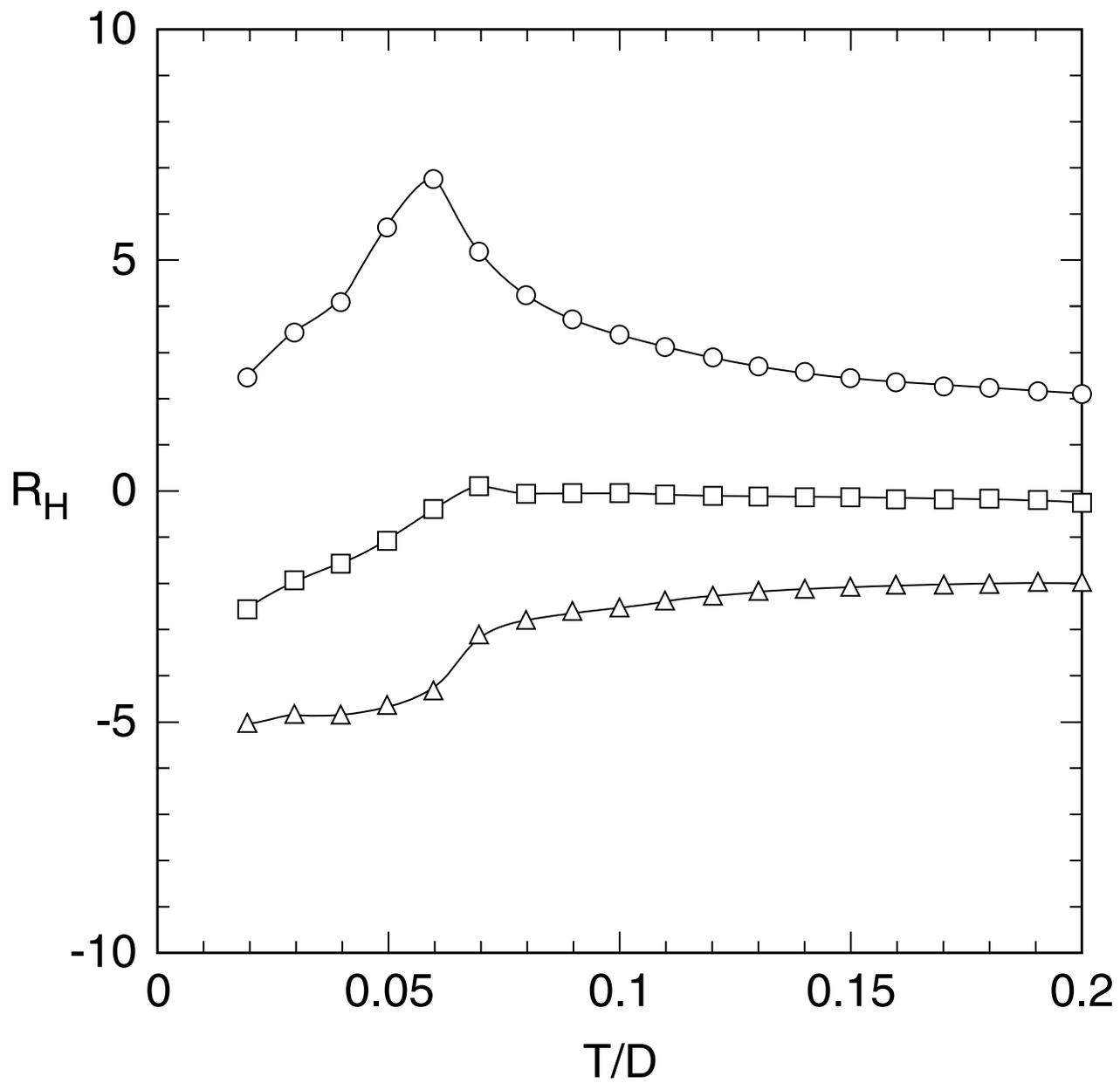
\end{document}